\documentclass[useAMS,usenatbib,fleqn]{mn2e}
\usepackage{amsmath,times}
\newcommand{\msol}{\rm M_\odot}
\input psfig

\title[Quiescent dwarf novae]
      {The behaviour of quiescent accretion discs in dwarf novae}
\author[M. R. Truss et al.]
       {M. R. Truss$^{1}$, G.A. Wynn$^{2}$ and P.J. Wheatley$^{2}$\\
\\
$^{1}$ School of Physics and Astronomy, University of St Andrews, North Haugh, St Andrews, Fife, KY16 9SS, Scotland, UK\\
$^{2}$ Department of Physics and Astronomy, University of Leicester, University Road, Leicester, LE1 7RH, UK}
 	
\begin{document}

\maketitle

\begin{abstract}
We propose a simple explanation for the constant mean brightness observed between outbursts in dwarf novae. Secular
changes in the total energy dissipation rate of the accretion disc brought about by variations in surface density, temperature 
and disc radius can be regulated by the gradual cooling of a small, critically-stable hot inner region. The hypothesis is supported with 
two-dimensional time-dependent numerical models of dwarf nova accretion discs. 
\end{abstract}


\section{Introduction}
Dwarf novae have been observed extensively since the discovery of U Geminorum in the mid-nineteenth century.
They account for well over half of the non-magnetic cataclysmic variables \citep{b15}, and consist of a white dwarf accreting
from a low-mass star via Roche lobe overflow. The dwarf novae are observed in two states: a quiescent, almost 
constant-luminosity low state that may last for months, and a regular (but aperiodic) high luminosity, outburst state lasting for
several days. The transition between the states is fairly rapid; typically the V magnitude changes by four magnitudes over the
course of a day or so.

The outbursts are attributed to a thermal-viscous instability in the accretion disc, often referred to as the disc instability, which 
is associated with the steep dependence of opacity with temperature at the point where hydrogen begins to ionize 
($T_{\rmn H} \sim$ 6500 K). Our understanding of the physical properties of the accretion disc in quiescence is less complete. In
particular, while great strides have been made in the identification of a magneto-rotational instability (MRI) responsible for the transport of
angular momentum \citep{b0}, the efficiency of this instability is sensitive to the degree of ionization in the accretion disc \citep{b21}. It  
remains unclear whether the MRI can sustain the necessary angular momentum transport appropriate to a cool, quiescent disc. It is 
possible that other viscosity mechanisms are at work in quiescence \citep{b13}.

There have been several rigorous one-dimensional calculations of dwarf nova outbursts within the framework of the
disc instability model (\citet{b9,b2,b12} and many others). The recent review of \citet{b5} describes the current state 
of the disc instability model, and contains an excellent commentary on its scope and limitations. In his review, Lasota
describes quiescence as 'the Achilles' heel' of the model, because the model handles the behaviour of dwarf novae in 
quiescence rather poorly. One of the major problems with the numerical models has been that when the correct boundary 
conditions are used (in particular an outer disc radius that is not constrained to be fixed) the models are unable to reproduce the
observed constant level of brightness during quiescence. The brightness almost always increases between the outbursts. 

Visual observations of dwarf novae by amateur astronomers cover more than a century in some cases and decades in many others 
(for example \citet{b4}). These visual observations show scatter at the level of 0.5 to 1 magnitudes, so it is difficult to see short-term 
variations, but it is clear that the mean visible flux remains approximately constant throughout quiescence. In contrast, disc instability 
models consistently predict an increasing brightness during quiescence. The amplitude of the predicted increase lies in the range 1 to 3 
magnitudes, and would certainly have been observed if present.

In this paper we suggest an explanation for this discrepancy. In the next section we present an analytic model of how the viscous energy 
generation rate responds to variations in the physical state of the disc over the quiescent interval in a dwarf nova. We show that in a disc with 
a constant Shakura-Sunyaev viscosity, the total viscous energy dissipation rate increases in response to the expected changes in surface
density, temperature and outer disc radius. However, we show that it is possible for the total viscous energy generation rate to remain 
constant if there is a small portion of the disc that remains in the high-viscosity state during quiescence. This hypothesis is 
supported by two-dimensional smoothed particle hydrodynamics (SPH) simulations of accretion discs in Section 3, which 
incorporate the evolution of the free outer disc boundary in response to the full two-dimensional tidal potential of the binary. We 
discuss the possible implications for the disc instability model in Section 4.

\section{Quiescent luminosity}

We consider the local viscous energy dissipation rate in an accretion disc, which is given by
\begin{equation}
D(r,t) = \frac{9}{4} \nu(t) \Sigma(r,t) \frac{GM_1}{r(t)^3}.
\label{diss}
\end{equation}
This definition differs by a factor of two from that which appears in some treatments. Here we express the local dissipation rate
integrated over the vertical interval $-\infty \le z \le \infty$, not just from the mid-plane. We assume that the viscosity can be 
described by the $\alpha$ prescription of \citet{b8}:
\begin{equation}
\nu = \alpha c_{\rmn {s}} H = \alpha c_s^2 \left(\frac{r^3}{GM_1}\right)^{\frac{1}{2}},
\label{ss}
\end{equation}
where the viscosity parameter $\alpha$ is a dimensionless constant.

Now, at a time $t$, the total luminosity due to viscous energy dissipation in the accretion disc is
\begin{equation}
L(t) = \int_{R_{\rmn{in}}}^{R_{\rmn{d}}} D(r,t) ~ 2 \pi r \rmn{d}r,
\label{dtot1}
\end{equation}
where $R_{\rmn{in}}$ and $R_{\rmn{d}}$ are the inner and outer radii of the disc. Then,
\begin{equation}
L(t) \propto \int_{R_{\rmn{in}}}^{R_{\rmn{d}}} \alpha T(r,t) \Sigma(r,t) r(t)^{-\frac{1}{2}} \rmn{d}r,
\label{dtot2}
\end{equation}
where $T(r,t)$ is the local mid-plane temperature at time $t$.

Since the conditions in the accretion disc are not steady during quiescence, there is no exact analytic expression for the
time-dependent evolution of the surface density, $\Sigma(r,t)$ and the mid-plane temperature, $T(r,t)$. However, we do 
know that these quantities must increase during quiescence if another outburst is to follow via disc instability. Indeed,
given that the critical surface densities for triggering and suppressing the disc instability vary almost linearly with radius \citep{b3},
one could envisage a model in which the surface density profile remains linear with radius while its gradient or mean level steadily 
increases. Further, we could assume that the radial temperature profile remains almost flat during quiescence, as suggested by 
eclipse mapping of the dwarf nova Z Cha \citep{b14}. We do not make these assumptions, but for simplicity we express the
product
\begin{equation}
T(r,t) \Sigma(r,t) = A(t) r^n
\label{adef}
\end{equation}
where $A$ is independent of radius and $n$ can take any value. Under the two assumptions discussed above, $n$ would be unity, 
but we keep it as a free parameter to take into account other possibilities. Its exact value (positive or negative) does not
affect our argument here, but we do assume that since temperature and surface density increase during quiescence, the quantity 
$\rmn{d}A/\rmn{d}t > 0$. In this case,
\begin{equation}
L(t) \propto \alpha_{\rmn{c}} A(t) \left[ R_{\rmn{d}}^{n+\frac{1}{2}} - R_{\rmn{in}}^{n+\frac{1}{2}}\right],
\label{dtgen}
\end{equation}
where the entire disc is on the cool branch of the disc instability curve, with a viscosity parameter $\alpha_{\rmn{c}}$. 
If we consider how the luminosity will vary during quiescence, assuming that the inner radius remains constant,
\begin{multline}
\frac{\rmn{d}L}{\rmn{d}t} \propto \alpha_{\rmn{c}} \left[ \left(R_{\rmn{d}}^{n+\frac{1}{2}} - R_{\rmn{in}}^{n+\frac{1}{2}} \right) \frac{\rmn{d}A}{\rmn{d}t}\right.\\
+ \left.\left(n+\frac{1}{2}\right) A R_{\rmn{d}}^{n-\frac{1}{2}} \frac{\rmn{d}R_{\rmn{d}}}{\rmn{d}t} \right].
\label{dt1}
\end{multline} 
The disc radius is expected to decrease slowly between dwarf nova outbursts, since the accretion rate onto the primary is small 
and material with low specific angular momentum is accumulating near the outer edge from the gas stream. This has been 
confirmed by one-dimensional disc instability calculations \citep{b1,b2}. However, this term is dominated by the increasing 
surface density and temperature, and the total disc luminosity increases between outbursts.

Now consider a disc with two regions: a hot, inner region on the upper branch of the instability curve with $\alpha_{\rmn{h}}$, 
and a cool outer region as before. We term the transition radius between the regions $R_{\rmn{t}}$. In this case,
\begin{equation}
L(t) \propto \int_{R_{\rmn{in}}}^{R_{\rmn{t}}} \alpha_{\rmn{h}} A_{\rmn{h}}(t) r^{n-\frac{1}{2}} \rmn{d}r + \int_{R_{\rmn{t}}}^{R_{\rmn{d}}}  \alpha_{\rmn{c}} A_{\rmn{c}}(t)   r^{n-\frac{1}{2}} \rmn{d}r,
\label{dthotgen}
\end{equation}
or
\begin{multline}
L(t) \propto \left[ \left( \alpha_{\rmn{h}}A_{\rmn{h}} - \alpha_{\rmn{c}}A_{\rmn{c}} \right) R_{\rmn{t}}^{n+\frac{1}{2}} \right.\\
\left. + \alpha_{\rmn{c}}A_{\rmn{c}}R_{\rmn{d}}^{n+\frac{1}{2}} - \alpha_{\rmn{h}} A_{\rmn{h}} R_{\rmn{in}}^{n+\frac{1}{2}}\right].
\end{multline}
Now,
\begin{multline}
\frac{\rmn{d}L}{\rmn{d}t} \propto \left[ \left(n +\frac{1}{2} \right) \left(  \alpha_{\rmn{h}} A_{\rmn{h}} - \alpha_{\rmn{c}} A_{\rmn{c}} \right)R_{\rmn{t}}^{n-\frac{1}{2}} \frac{\rmn{d}R_{\rmn{t}}}{\rmn{d}t}\right.\\
~~~~~~~~~~~~~~~+\left(n+\frac{1}{2}\right) \alpha_{\rmn{c}} A_{\rmn{c}} R_{\rmn{d}}^{n-\frac{1}{2}} \frac{\rmn{d}R_{\rmn{d}}}{\rmn{d}t}\\
~~~~~~~~~~~~~~~~~~~~~~~~+~\alpha_{\rmn{h}} \left( R_{\rmn{t}}^{n+\frac{1}{2}} - R_{\rmn{in}}^{n+\frac{1}{2}} \right)\frac{\rmn{d}A_{\rmn{h}}}{\rmn{d}t}\\
\left.+\alpha_{\rmn{c}}\left(R_{\rmn{d}}^{n+\frac{1}{2}} - R_{\rmn{t}}^{n+\frac{1}{2}}\right)\frac{\rmn{d}A_{\rmn{c}}}{\rmn{d}t}\right].
\label{long}
\end{multline}
The second and fourth terms of this equation are identical to the two terms in equation \ref{dt1}; however, there are now two
additional terms involving the rate of change in $R_{\rmn{t}}$ and $A_{\rmn{h}}$. The presence of a hot inner region means that
there are more parameters available to change $\rmn{d}L / \rmn{d}t$. As an example, the coefficients of the two new terms in 
equation \ref{long} involve the hot-state viscosity, $\alpha_{\rmn{h}}$, while the coefficients of the other terms involve only 
$\alpha_{\rmn{c}}$. Since we expect $\alpha_{\rmn{h}} \sim 10 \alpha_{\rmn{c}}$ in a dwarf nova accretion disc, any changes in 
dissipation rate brought about by the evolution of the cool part of the disc could be offset by much smaller changes in the hot, 
inner region. 

We stress that the above analysis is in no way meant to be a self-consistent solution of the accretion flow. We simply
demonstrate that a hot, high-viscosity inner region provides many more avenues for evolution of the disc. For example, the 
situation is complicated by the fact that we do not know the correct value of $n$ or the correct form of $A(t)$. In particular, it is 
extremely difficult to predict the response of $R_{d}$ and $R_{t}$ in a non-steady accretion disc. It is only in a numerical calculation 
that these effects can be studied in more detail.

In the next section, we demonstrate the effect of maintaining a hot, high-viscosity inner region with two-dimensional smoothed 
particle hydrodynamics calculations of accretion discs in dwarf novae.

\section{Two-dimensional SPH calculations}
\subsection{Numerical method}

We use a two-dimensional (2D) smoothed particle hydrodynamics scheme to model the evolution of an accretion disc in a close
binary. The model includes the full tidal potential due to the secondary star and has been discussed in detail and tested with reference 
to dwarf novae in \citet{b10}. 

SPH uses an ensemble of particles to describe a fluid. The advantages of such a Lagrangian scheme to model these systems 
immediately becomes apparent: the response of the free outer boundary of the accretion disc to the mass stream and the tidal potential 
are treated in a self-consistent manner. For a wider discussion of SPH, the reader is referred to the review of \citet{b6}. 

We solve the SPH momentum equation in the Lagrangian (co-moving) frame, with an artificial viscosity parameter $\zeta$ for each 
particle, $i$, such that
\begin{equation}
\frac{\rmn{d}{\bf v}_i}{dt} = -\sum_jm_j \left( \frac{P_i}{\rho_i^2}+\frac{P_j}{\rho_j^2}-\frac{\zeta\bar c_{ij}\mu_{ij}}{\bar \rho_{ij}} \right) \nabla_i W_{ij}
\label{mom}
\end{equation}
where $\bar \rho_{ij}$ and $\bar c_{ij}$ are the average density and sound speed of a pair of particles $i$ and $j$ respectively, 
$W_{ij}$ is an interpolating kernel (here taken to be a cubic spline) and 
\begin{equation}
\rm  \mu_{ij} = \frac{H{\bf v}_{ij} \cdot {\bf r}_{ij}}{{\bf r}_{ij}^2 + \eta^2}.
\label{piv}
\end{equation}
The softening parameter $\eta$ avoids singularities and H is the length scale over which the viscous energy is dissipated. Here,
this is set to the scale height, $H = \frac{c_{\rmn{s}}}{\Omega}$. We do not use a viscosity term that is quadratic in $\mu_{\rmn{ij}}$, 
often denoted by the parameter $\beta$ in SPH equations. A
linear term is quite sufficient in this case. \citet{b7} has shown that in 2D, the $\zeta$ term introduces a shear viscosity
\begin{equation}
\nu = \frac{1}{8} \zeta c_{\rmn{s}} H,
\end{equation}
so the implementation is equivalent to a Shakura-Sunyaev viscosity of $\alpha = \zeta /8$.
We also solve the thermal energy equation
\begin{equation}
\frac{\rmn{d}u_i}{\rmn{d}t} = \frac{1}{2}\sum_{j=1}^N m_j \left[\frac{P_i}{\rho_i^2} + \frac{P_j}{\rho_j^2} - \frac{\zeta\mu_{ij}\bar c_{ij}}{\bar \rho_{ij}} \right] ({\bf v}_i - {\bf v}_j) \cdot {\bf \nabla}_i W_{ij}.
\label{thermen}
\end{equation}
\noindent

The thermal-viscous instability is incorporated using a viscosity switch for each particle based on the local surface density. The
critical conditions for the triggering or suppression of the instability are roughly linear in radius \citep{b3}. Here, we take
\begin{equation}
\Sigma_{\rmn{max}} = C_{\rmn{hot}} \frac{R}{a} ~~~;~~~ \Sigma_{\rmn{min}} = C_{\rmn{cold}} \frac{R}{a},
\label{crit}
\end{equation}
where $ C_{\rmn{hot}}$ and $C_{\rmn{cold}}$ are constants and $a$ is the binary separation. If the surface density of a particle exceeds
$\Sigma_{\rmn{max}}$, then its viscosity is increased smoothly from a value $\zeta_{\rmn{c}}$ to a value $\zeta_{\rmn{h}}$ on a 'thermal' 
time-scale $\tau_{\rmn{trigger}}$, where  $\tau_{\rmn{trigger}} << t_{\rmn{viscous}}$. Its viscosity remains high until the local surface 
density falls below $\Sigma_{\rmn{min}}$, when it is returned to $\zeta_{\rmn{c}}$ on a similar time-scale. In contrast to previous
implementations of this code which were fully isothermal, here we change the sound speed of the gas whenever the critical surface density
thresholds are crossed. The change in sound speed from a value $c_{\rmn{low}}$ to a value $c_{\rmn{high}}$ (and vice versa) is
made on precisely the same time-scale as the change in viscosity parameter. Hence, no scaling is required for the amplitude of the
luminosity variations driven by viscous dissipation in the disc.

\subsection{Simulations}

We perform simulations of an accretion disc in a binary stellar system with the parameters of the dwarf nova IP Pegasi taken from the 
catalogue of \citet{b15}: 
$P_{\rmn{orb}} = 0.158$ d, $M_1 = 1.15 \rmn{\msol}$ and $M_2 = 0.67 \rmn{\msol}$. The accretion disc is built up from scratch using a
stream of particles injected from the inner Lagrangian point at a constant rate of $10^{-10} ~ \rmn{\msol yr^{-1}}$. We use 
$C_{\rmn{hot}} = 18.7~\rmn{g cm^{-2}}$, $C_{\rmn{cold}} = 7.02~\rmn{g cm^{-2}}$, $\tau_{\rmn{trigger}} =  1750~s$, viscosity 
parameters $\zeta_{\rmn{c}}$ = 1 and $\zeta_{\rmn{h}}$ = 10, corresponding to $\alpha_{\rmn{c}}$ = 0.125 and $\alpha_{\rmn{h}}$ = 1.25, and
sound speeds $c_{\rmn{low}} = 0.05 a\Omega_{\rmn{b}}$ and $c_{\rmn{high}} = 0.15 a\Omega_{\rmn{b}}$.
\begin{figure}
\psfig{file=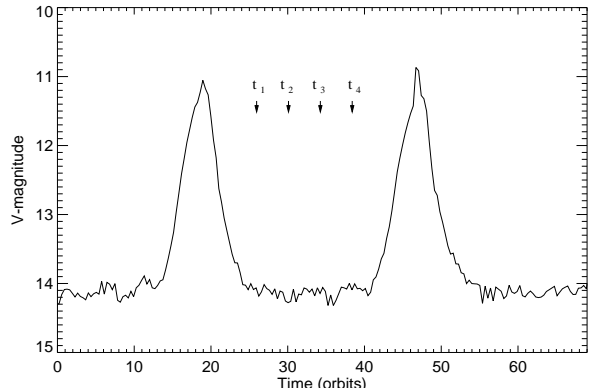,width=8cm}
\caption{V-band light curve for a simulation of the dwarf nova IP Peg. Here, the inner part of
the disc remains in the hot, high viscosity state throughout the quiescent intervals. The size of the hot region varies during 
quiescence, leading to the small variations in the light curve, but keeping the mean quiescent level flat. The four times $t_1$ to $t_4$ are
defined in the caption of Figure 2.}
\label{flat}
\end{figure}
\begin{figure}
\psfig{file=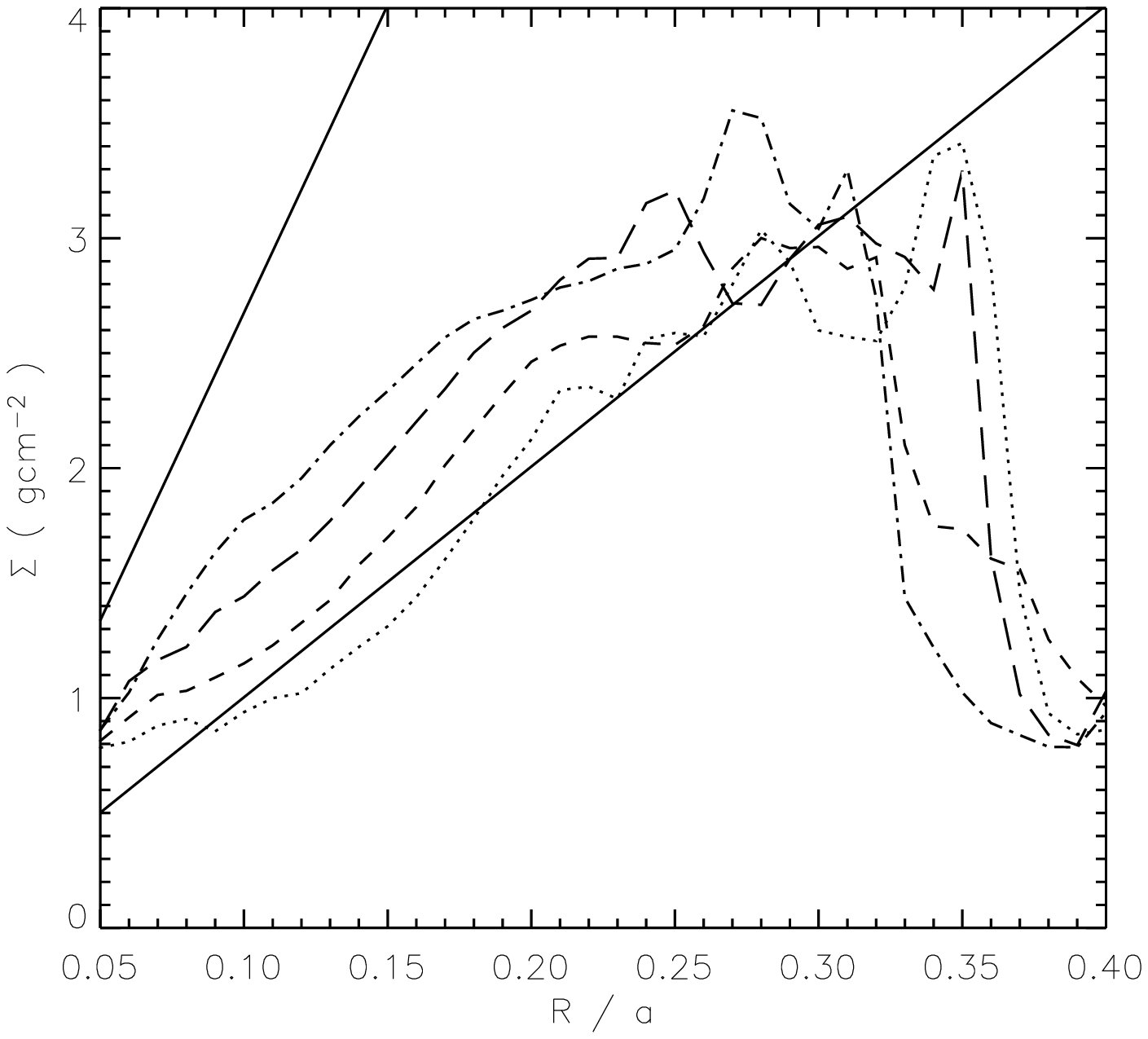,width=8cm}
\caption{Evolution of surface density in the accretion disc during quiescence in the first simulation. The curves 
correspond to the following times in Figure \ref{flat}: $t_{\rmn{1}}$=26 (dots), $t_{\rmn{2}}$=30 (dashes), $t_{\rmn{3}}$=34 (long dashes) 
and $t_{\rmn{4}}$=38 (dot-dashes). The solid straight lines show the critical surface densities used in the simulation.}
\label{sig}
\end{figure}
\begin{figure*}
~~~\psfig{file=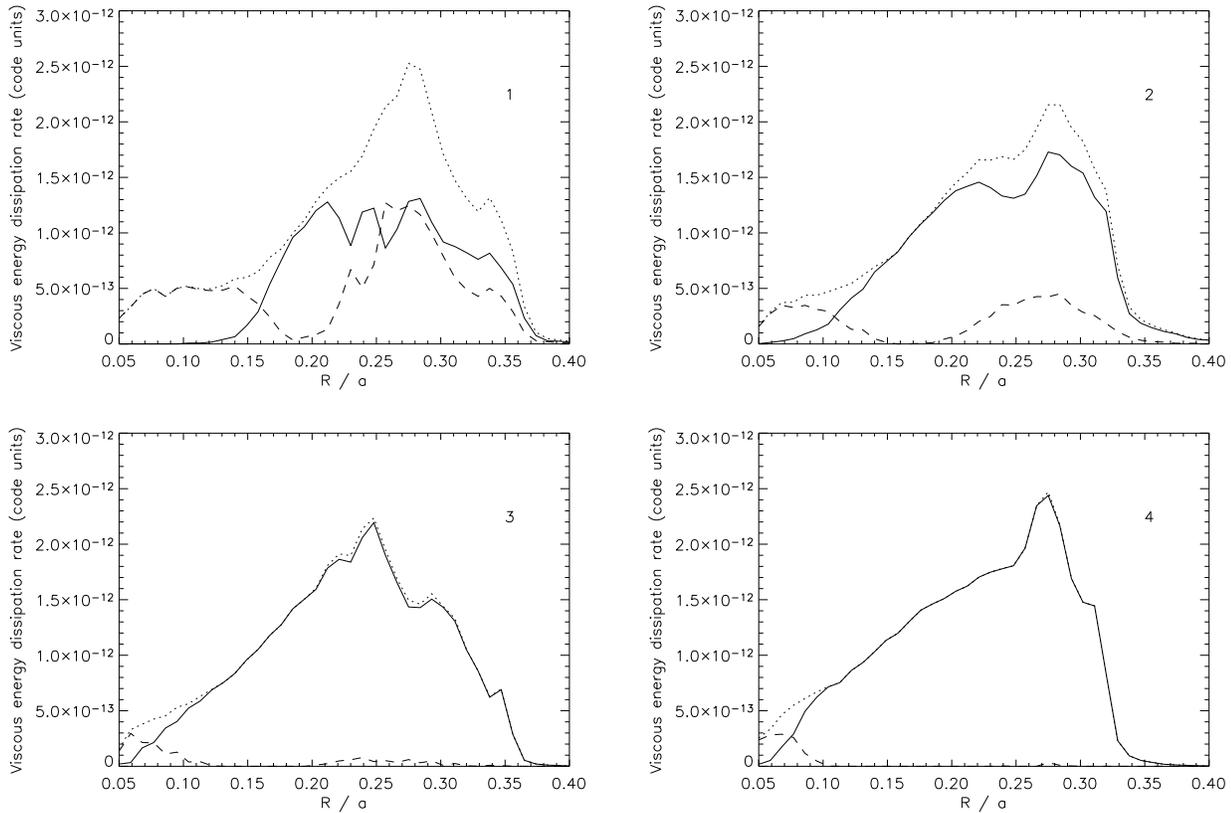,width=17cm}
\caption{Evolution of the contributions to the viscous energy dissipation rate made by the high and low viscosity particles in the disc for
each of the times $t_{\rmn{1}}$ (top left) to $t_{\rmn{t4}}$ (lower right). The solid curve shows the contribution from low viscosity 
particles ($\zeta \leq 1.5$), the dashed curve shows the contribution from high viscosity particles ($\zeta > 1.5$), and the dotted curve 
the total viscous energy dissipation rate. The increasing low-viscosity contribution is offset by a decrease in the contribution from the
high viscosity particles.}
\label{hotcold}
\end{figure*}

Figure \ref{flat} shows the resulting V-band light curve. The light curve is calculated under the assumption that the disc is optically thick and
each annulus of gas radiates as a black body. The Planck function is integrated over the interval 500 to 600 nm and summed over the annuli. 
In this simulation, many particles in the inner part of the disc remain constantly in the hot, high-viscosity state. In this region 
the critical surface densities given in the equations \ref{crit} are both low and close together. The overall mean brightness is 
roughly constant during quiescence, with small variations occurring as the size of the hot region changes. It can be seen that this behaviour is 
now consistent with that observed in dwarf in quiescence. Our predicted variability in quiescence has too small an amplitude to be detected 
the visual light curves of amateur astronomers, but it is strikingly similar to that observed in the few detailed CCD observations of dwarf novae
in quiescence (see, for example \citet{b17}).

Figure \ref{sig} shows the variation in surface density during one of the quiescent intervals. It can be seen that the surface density in the inner
disc never drops below $\Sigma_{min}$ and hence remains in the hot state throughout quiescence.

A hot inner disc seems to be a natural feature of these simulations, and in fact is a natural consequence of the linear triggers that get 
progressively closer together in the inner disc. With triggers of this form it is difficult to see how the inner regions of the disc can avoid entering 
the high state during quiescence. It is not immediately obvious to us why this is not a feature seen in one-dimensional models.

The response of the disc is shown in more detail in Figure \ref{hotcold}, which shows the evolving contributions to the local viscous energy 
dissipation rate of the low and high viscosity particles at each radius in the disc. For the purposes of this figure we define the high viscosity particles 
as those with $\zeta > 1.5$, to distinguish between those particles with a genuinely low viscosity, such as those arriving from the stream which all 
have $\zeta$ = 1.0, and those which may be changing viscosity from a recent high state. At the beginning of quiescence, when the surface density and 
dissipation rate in the outer disc are low, the high viscosity particles dominate the inner disc. The hot particles at large radii ($R > 0.2a$) in panels 1 
and 2 remain from the previous outburst. By mid-quiescence (panel 3), they have cooled completely and the outer disc is dominated by the cold
particles arriving from the mass stream. As the cold, outer regions begin to refill, the corresponding increase in dissipation is offset by a 
decrease in the size of the residual high viscosity inner region. An estimate for the transition radius, $R_{\rmn{h}}$, can be made from the 
intersection of the solid and dashed curves in Figure \ref{hotcold}. Although $R_{\rmn{h}}$ is varying 
both up and down during quiescence, the general trend is a decrease from $R_{\rmn{h}} \sim 0.15a$ in panel 1, to $R_{\rmn{h}} \sim 0.08a$ in 
panel 4.

Throughout the quiescence, the outer radius shrinks slightly, in accordance with the expectations discussed in Section 2. We find that over the
course of quiescence, although the mean luminosity remains constant, the total disc mass increases by 19 \%, which is comparable to the mass
accreted by the primary in each outburst.

A second simulation of IP Peg was performed in order to demonstrate that it is the hot inner parts that keep the mean quiescent 
brightness constant. In this case, we used critical surface densities that did not have a linear radial dependence, but were 
constant: $\Sigma_{\rmn{max}} = C_{\rmn{up}}$ and $\Sigma_{\rmn{min}} = C_{\rmn{down}}$, where we used 
$C_{\rmn{up}} = 4.1~\rmn{gcm^{-2}}$ and  $C_{\rmn{down}} = 1.6~\rmn{gcm^{-2}}$. In this case, it is more unlikely that the surface density of 
the disc in the inner parts can reach the critical level during quiescence, so the inner part of the disc remains in the cool, low-viscosity state. 
\begin{figure}
\psfig{file=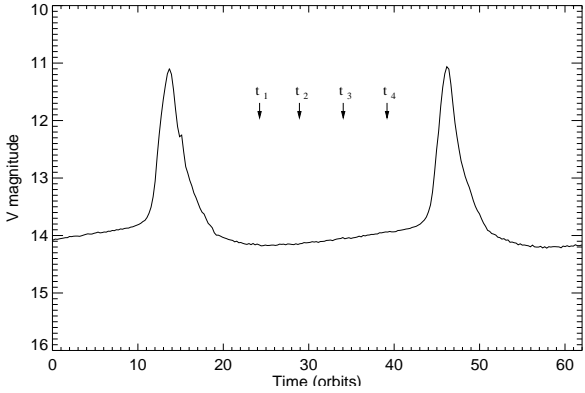,width=8cm}
\caption{In this simulation of outbursts in IP Peg, the surface density trigger for the disc instability in the inner part of the disc is 
raised so that this region cannot reach the hot, high viscosity state during quiescence. Consequently, the mean brightness level 
rises steadily between outbursts. The four times $t_1$ to $t_4$ are defined in the caption of Figure 6.}
\label{inc}
\end{figure}

\begin{figure}
\psfig{file=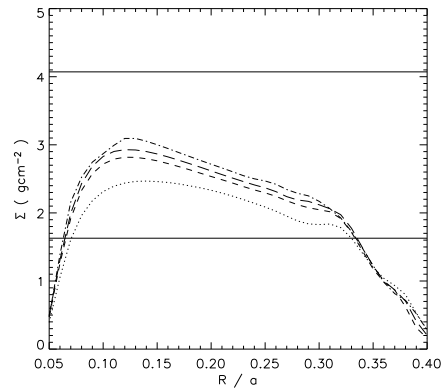,width=8cm}
\caption{Evolution of surface density in the accretion disc during quiescence in the second simulation. The curves 
correspond to the following times in Figure \ref{inc}: $t_{\rmn{1}}$=24 (dots), $t_{\rmn{2}}$=29 (dashes), $t_{\rmn{3}}$=34 (long dashes) and 
$t_{\rmn{4}}$=39 (dot-dashes). The solid straight lines show the critical surface densities used in the simulation.}
\label{sigflat}
\end{figure}

Figures \ref{inc} and \ref{sigflat} show the resulting V-band light curve and variations in surface density during the second simulation. 
The outbursts are almost identical in profile, and the more familiar solution is recovered, with a mean quiescent 
level that steadily increases between the outbursts as gas accumulates in the disc. 

The increase in V-magnitude during quiescence 
($\sim 0.2 - 0.3$ mag.) is smaller than the increase seen in 1-d models ($\sim 1$ mag.). This can be understood in terms of the sound speed, 
which only changes (that is, the disc only heats up) when the viscosity changes. In the 1-d models, the temperature and 
sound speed of the disc increase during quiescence as the surface density increases, even when $\alpha = \alpha_{\rmn{cold}}$ everywhere.
It is trivial to estimate the effect that this change in sound-speed would have on our simulation. The dissipation rate, $D(r) \sim \nu \Sigma \sim
T_{\rmn{c}} \Sigma $, where $T_{\rmn{c}}$ is the mid-plane temperature. In our simulation, the dissipation rate at the end of quiescence is
about 1.25 times (0.2 magnitudes larger than) the rate at the beginning of quiescence. This arises from the increase in surface density alone. 
Given a reasonable increase of a factor of two in $T_{\rmn{c}}$ over quiescence, the increase in dissipation rate would be a factor of two
higher, that is a difference of 2.5 times (or 1 magnitude). Indeed, we can go one step further and incorporate this into the simulations.
\begin{figure}
\psfig{file=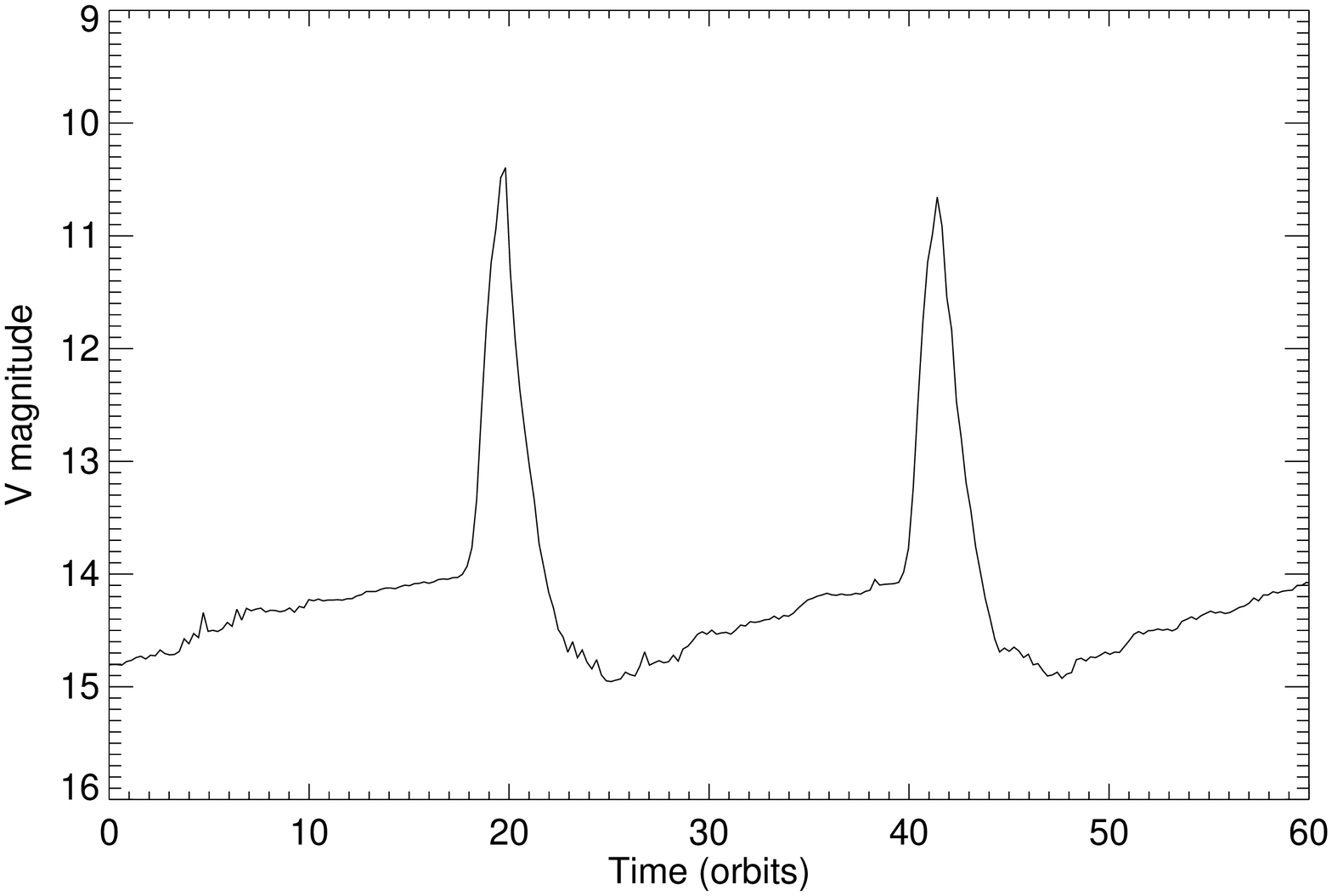,width=8cm}
\psfig{file=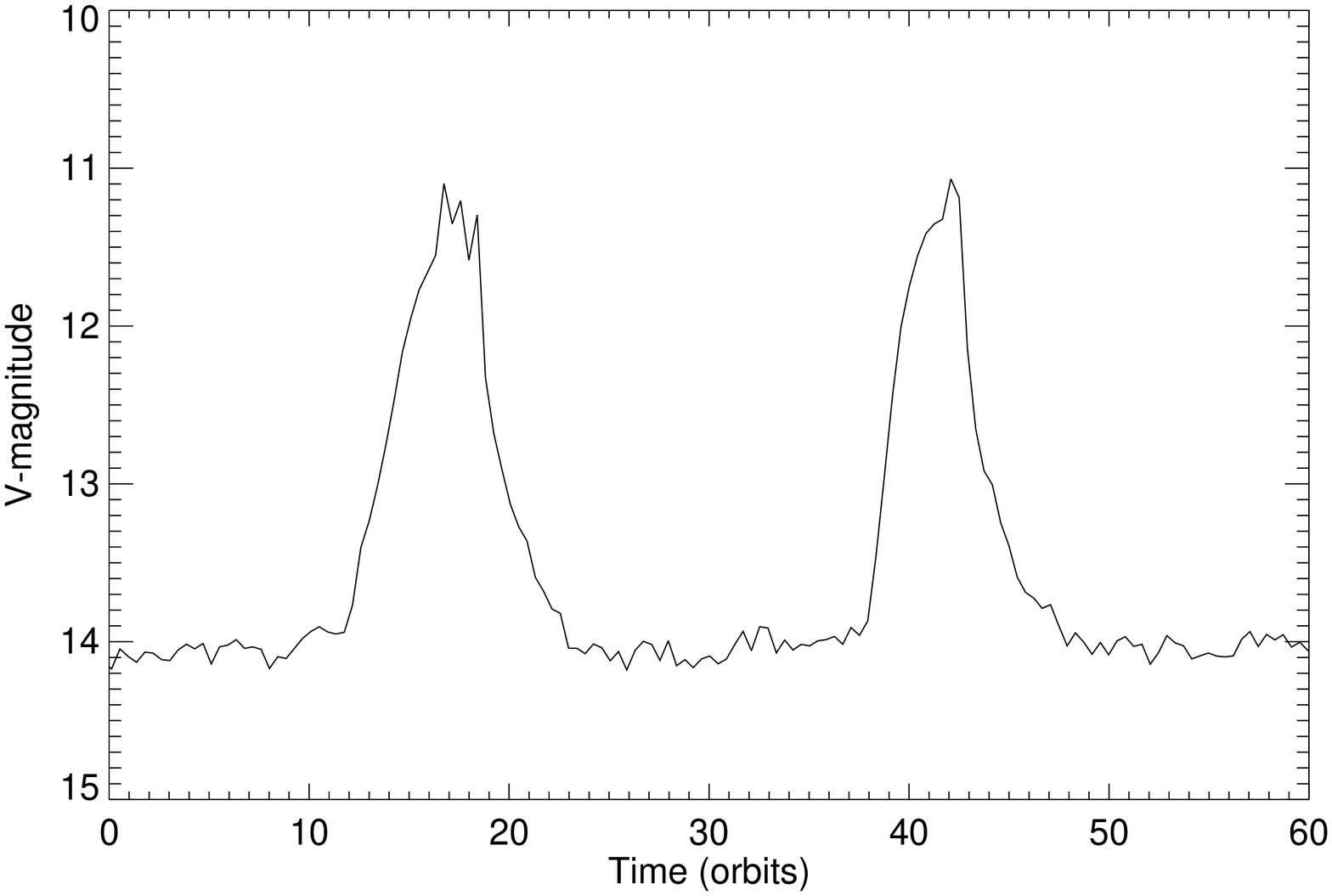,width=8cm}
\caption{Evolution of the V-band light curve when the sound speed is varied with surface density for the cold particles. Top: the light curve of the
second (flat-trigger) simulation now shows a brightening by a full magnitude in quiescence. Bottom: the difference in the first simulation is
much less marked.}
\label{newlc}
\end{figure}

In an $\rmn \alpha$-disc, surface density scales with sound speed as $\Sigma \sim c_{\rmn{s}}^{3/14}$. For all particles with
$\alpha = \alpha_{\rmn{c}}$, we let
\begin{equation}
c_{\rmn{s}} = c_{\rmn{low}} \left( \frac{\Sigma}{\Sigma_\rmn{min}} \right)^{\frac{3}{14}}.
\end{equation}
The results can be seen in Figure \ref{newlc}. For the flat-trigger case (simulation 2), where most of the particles during quiescence have
$\alpha = \alpha_{\rmn{cold}}$, the rise in the V-band light curve is now a full magnitude, in line with our expectations. The difference in 
simulation 1 is much less marked, because there the viscosity is not $\alpha_{\rmn{cold}}$ everywhere during quiescence. The amplitude of 
the variations in quiescence in the first simulation remain much smaller than the increase in V-band magnitude  in the second simulation.

\begin{figure}
\psfig{file=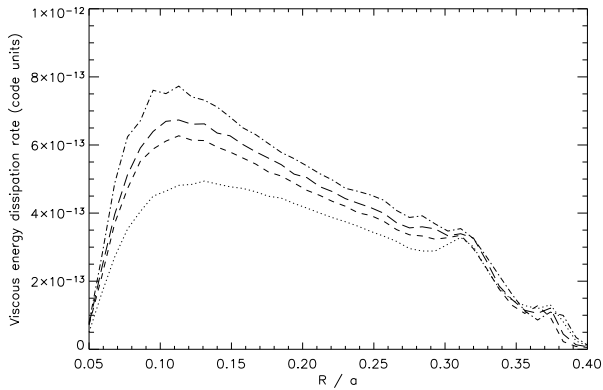,width=8cm}
\caption{Evolution of total viscous energy dissipation rate in the second simulation. The lines correspond to the same times as those in Figure 
6.}
\label{duflat}
\end{figure}

Figure \ref{duflat} shows the evolution of total viscous energy generation over the disc in the second simulation. Again, the outer radius moves 
inward slightly, but there is a general increase in dissipation rate at all radii. The increase in disc mass is virtually identical to that found in 
simulation 1, around 19 \%.

\section{Discussion}

We have shown that the observed constant mean brightness of dwarf novae during quiescence can be regulated by the response of a small inner 
portion of the disc that remains in a high-viscosity state close to the critical surface density limits for the thermal-viscous disc instability. Over the
course of quiescence, the increase in dissipation in the outer disc brought about by the increasing surface density and central
temperature is offset by the gradual cooling and shrinking of the high-viscosity inner region.

A steady increase in the mean brightness level  between outbursts is unavoidable if the entire disc is in a single viscosity state,
because both the surface density and temperature must be increasing. This has long been known to be the case from 
thermodynamically-detailed one-dimensional calculations in which the quiescent disc is entirely in the cool, low viscosity state. The
effect of increasing $\Sigma$ and $T$ dominates the effect of the disc shrinking in quiescence (equation \ref{dt1}), and the brightness
rises steadily. 

However, we can see from equation \ref{long} that this problem can be resolved if a small region remains in the high-viscosity state 
while the surface density and temperature increase in the rest of the disc. In the SPH calculations, this is achieved by the surface
density hovering near the critical stability limit at all times. The high-viscosity, hot inner disc is the main difference between our 
simulations and those within the framework of the standard disc instability model \citep{b5}, where the quiescent disc is cold everywhere and the 
surface density is left well below $\Sigma_{\rmn{min}}$ after each outburst.

The key question that will remain is how can this be achieved in a real accretion disc? It has been suggested that the inner parts of the accretion disc 
can be heated and evaporated near the white dwarf \citep{b16}. Indeed, a low density region in the inner disc can explain the UV delay 
observed in dwarf novae, because at the onset of an outburst it takes a few hours for the low density region to refill. Of course, the UV delay will be
present however the low density region arises: in addition to the evaporation model of \citet{b16}, \citet{b22} have suggested that the inner part of 
a quiescent disc can be evacuated by the weak magnetic field of the white dwarf primary. An attractive feature of the models that we have presented in 
this paper is that a similar low-density inner disc is predicted during quiescence. Here, however, the small low density region is a consequence of a local 
viscosity higher than that which is usually assumed for a quiescent disc. This is exactly analogous to the situation found by \citet{b23}, who showed that
the UV delay can be reproduced if irradiation by the hot white dwarf keeps the disc hot and fully ionised out to several stellar radii.

Although there are difficulties separating spectral components due to the accretion flow itself and the white dwarf in quiescent dwarf novae, there is 
observational evidence to support the presence of such a hot, optically thin inner region (see, for example, \citet{b18,b19}). Such a flow would almost 
certainly be highly turbulent, and it is exactly these conditions under which the MRI is expected to be most efficient and the viscous stress high. 

We conclude that the constant brightness level that is observed between the outbursts of dwarf novae is well explained by the
gradual cooling of a small, critically-stable, hot inner region. The diminishing contribution to the viscous energy dissipation rate
from such a region is sufficient to mask the increasing energy dissipation due to increases in mid-plane temperature and 
surface density in the rest of the cool disc.

\section*{Acknowledgments}

MRT acknowledges a PPARC Postdoctoral Fellowship.

\end{document}